\documentclass[preprint]{elsarticle}

\usepackage{lineno}
\usepackage[ruled,vlined]{algorithm2e}
\usepackage{graphics,amsmath,amsfonts,times,multirow,array,booktabs,caption,color}
\usepackage[toc,page]{appendix}
\usepackage{subfig}
\modulolinenumbers[5]

\journal{Journal of Computational Physics}

\begin{document}

\begin{frontmatter}

\title{An Approximate Framework for Quantum Transport Calculation with Model Order Reduction}


\author[mymainaddress]{Quan CHEN\corref{mycorrespondingauthor}}

\author[mysecondaryaddress]{Jun LI}

\author[mythirdaddress]{Chiyung YAM}

\author[mysecondaryaddress]{Yu ZHANG}

\author[mymainaddress]{Ngai WONG}

\author[mysecondaryaddress]{Guanhua CHEN}

\cortext[mycorrespondingauthor]{Corresponding author. Email: quanchen@eee.hku.hk. Tel: (852) 2219-4846}

\address[mymainaddress]{Department of Electrical and Electronic Engineering, The University of Hong Kong, Hong Kong}
\address[mysecondaryaddress]{Department of Chemistry, The University of Hong Kong, Hong Kong}
\address[mythirdaddress]{Beijing Computational Science Research Center, China}

\begin{abstract}
A new approximate computational framework is proposed for computing the non-equilibrium charge density in the context of the non-equilibrium Green's function (NEGF) method for quantum mechanical transport problems. The framework consists of a new formulation, called the X-formulation, for single-energy density calculation based on the solution of sparse linear systems, and a projection-based nonlinear model order reduction (MOR) approach to address the large number of energy points required for large applied biases. The advantages of the new methods are confirmed by numerical experiments.
\end{abstract}


\end{frontmatter}

\linenumbers

\section{Introduction}
With the advent of $10$nm technology node, quantum mechanical (QM) phenomena have emerged as a central issue in modeling and simulation of nano-devices wherein numerical calculations based on first-principle QM physics have become indispensable. The non-equilibrium Green's function (NEGF) method~\cite{Datta:00,Datta:02} is the most widely used numerical tool for QM transport problems, providing an effective approach to solve the Schr\"odinger equation self-consistently with the Poisson equation. The most computation intensive step in NEGF is the calculation of charge density from the Green's function, which involves computing a subset of entries in the inverse of a large sparse matrix and has to be repeated many times. In this context, the recursive Green's function (RGF) method~\cite{Haydock:72,Sols:89} has been the workhorse over years. With a mathematical origin in the semi-separable matrix~\cite{Jitesh:07}, RGF allows efficient calculation of some blocks in the inverse of a block tri-diagonal matrix without forming the whole inverse. While effective for long and thin structures, RGF becomes inefficient when the target structures have a large cross-section. The complexity of RGF can be estimated as $O(N_x^3N_y)$, where $N_x$ is the average size of the matrix blocks treated by dense matrix algebra and is proportional to the layer size, and $N_y$ is the number of the blocks in scale with the length. Therefore, the cubic growth of computation with the cross-sectional area renders RGF difficult to be used to simulate realistic devices wherein the cross-section of current-carrying channel easily exceeds $10\times 10 nm^2$. To mitigate the performance bottleneck, efficient alternatives have been proposed, such as the FIND method~\cite{Li:08,Li:12} based on the nested dissection and the selected inversion (SelInv)~\cite{Lin:09,Lin:11} based on the hierarchical Schur complement. All of these methods exhibit a lower asymptotic complexity than RGF by exploiting more advanced matrix partitioning techniques than the 1D partitioning used by RGF~\cite{Hetmaniuk:13}.

Another challenge facing NEGF in nano-device modeling lies in the large amount of energy points needed to evaluate the non-equilibrium (NEQ) charge density, due to the highly oscillatory integration along the real energy axis. Typical values used in practice are $400-500$ points per eV~\cite{Guo:13}. Simulations involving large devices, as a consequence, are usually performed under small applied bias, e.g., $50mV$, so as to limit the number of energy points to a few dozen. However, realistic biases are in the range of $0.5V \sim 1V$, and $10$ times more energy points as well as computation are thereby demanded. On the other hand, since these energy points are "clustered" within a relatively small range, it is reasonable to expect that the solutions at these points are not truly independent of each other; they may have substantial overlap in the information they carry. Such information correlation has not been fully exploited by existing methods, in which each energy is solved individually.

Model order reduction (MOR) represents a family of mathematical techniques aiming to reduce the number of degrees of freedoms in a numerical system, while preserving important system properties such as I/O relations~\cite{Schilders:08}. The rationale underlying MOR is that, with efforts spent in building a smaller system that to some extent mimics the original large system, significant speedup can be achieved by simulating this reduced-order-model (ROM), and recovering afterwards the responses of the original system from that of the ROM. Further saving is available if the ROM can be used repeatedly, e.g., in a design exploration phase. Originated from system control theory, MOR (or equivalents under different names) has been found numerous applications in many different scientific areas. Given the large number of repetitions of charge density evaluation in NEGF, MOR comes as a natural option to bring down the computation cost. However, most existing MOR techniques require the system of interest to be described in the state-space or descriptor representations~\cite{Tan:07}. The present formulation of NEQ charge density is not in these desired forms, and this hinders the application of MOR.

In this paper, we aim to develop an approximate computation framework for the NEQ calculation in NEGF. The method consists of two ingredients. The first ingredient is a new formulation of the single-energy density calculation, based on solving a sparse linear system with multiple right-hand-side (RHS) vectors. Different from RGF, FIND or SelInv which are all direct methods, the proposed formulation, dubbed X-formulation since it is based on solving $AX=b$, allows tradeoff between accuracy and computational cost. It also enables the use of iterative methods, a well-established field with abundant literature, to replace the direct LU (or LDLT) factorization for long-term scalability. More importantly, the X-formulation removes the obstacle to applying MOR by transforming the NEQ charge density calculation problem into a state-space representation. Hence, a MOR technique is developed to be the second ingredient of our framework. Because of the nonlinear nature of the problem, we choose a projection-based MOR method based on the parameter-dependent Krylov subspace (PDKS). The idea is to solve the full-size problem at a small number of energy points, use the solutions to construct a subspace basis, and perform the numerical quadratures with the ROM obtained by projecting the original system onto this subspace.

The rest of the paper is organized as follows. Section 2 will review the problem formulation for NEQ charge density calculation and the exact methods. Section 3 will describe the new X-formulation for single-energy calculation. Section 4 will give the details of the MOR algorithm. Section 5 will present numerical results and Section 6 will draw the conclusion.

\section{Non-Equilibrium Charge Calculation in NEGF}
For simplicity, we assume a two-terminal device and do not include scattering mechanisms throughout the paper. The central quantity in NEGF is the (retarded) Green's function $G(E) \in \mathbb{C}^{N\times N}$ with $N$ being the number of orbitals, which reads
\begin{equation}\label{eq:GF}
    G(E) = M(E)^{-1} = \left[ES-H-\Sigma_L(E)-\Sigma_R(E)\right]^{-1},
\end{equation}
where $E$ is the energy (scalar), $S\in \mathbb{R}^{N\times N}$ is the overlap matrix (real symmetric) for non-orthogonal basis, $H \in \mathbb{R}^{N\times N}$ the Hamiltonian (real symmetric but indefinite) of the device under consideration and $\Sigma_{L,R}(E) \in \mathbb{R}^{N\times N}$ the self-energy matrices (complex symmetric) accounting for the influence of the left and right leads. Within the NEGF formalism $\Sigma_L(E)$ and $\Sigma_R(E)$ are nonlinear functions of $E$ without close-form expression, and usually have only small $N_c \times N_c$ nonzero blocks at the top-left and the lower-right corner, respectively. Let $\mathcal{I}_L$ and $\mathcal{I}_R$ be two integer vectors containing the matrix indices corresponding to the left and right leads respectively, then the nonzero blocks of the self-energy matrices are given by
\begin{align}\label{eq:Sigmablk}
    \Sigma _L^{blk} &= {\Sigma _L}\left( {{\mathcal{I}_L},{\mathcal{I}_L}} \right) = {\tau _{DL}}{g_{LL}}\tau _{DL}^\dag \\ \nonumber
    \Sigma _R^{blk} &= {\Sigma_R}\left( {{\mathcal{I}_R},{\mathcal{I}_R}} \right) = {\tau _{DR}}{g_{RR}}\tau _{DR}^\dag
\end{align}
where $\tau_{DL} = Es_{DL}-h_{DL}$ and $\tau_{DR} = Es_{DR}-h_{DR}$ are the outer blocks coupling the leads and the device, and $g_{LL}$ and $g_{RR}$ are the surface Green's functions of the left and right leads.

The NEQ Mulliken charge of each orbital is defined by the energy integral
\begin{equation}\label{eq:Q}
Q_{neq} = -\frac{2}{\pi}\int_{{E_b}}^{{E_t}} {dE\; q(E)}
\end{equation}
where $E_b$ and $E_t$ are respectively the bottom and top energy levels for the integral. The single-energy electron density consists of three parts
\begin{equation}\label{eq:q}
    q(E) = diag\{G^<(E) S\} - diag\left\{ {{\widetilde G^<_{DLt}}{{\left( E \right)} }{s_{DL}^T}} \right\} - diag\left\{ {{\widetilde G^<_{DR}}\left( E \right)s_{DR}^T } \right\}
\end{equation}
where $G^<(E)$ is the lesser Green's function of the device defined by (assuming higher potential in the right lead)
\begin{equation}\label{eq:Gless}
     G^< \left( E \right) = G\left( E \right){\Gamma _R}\left( E \right) G\left( E \right)^\dag
\end{equation}
with $\Gamma _R = \operatorname{Im}(\Sigma_R)$. The last two terms in (\ref{eq:q}) involve the outer blocks of the lesser Green's function, $\widetilde{G}^<_{DL},\widetilde G^<_{DR} \in \mathbb{C}^{N\times N_c}$ and the outer blocks of the overlap matrix, $s_{DL},s_{DR} \in \mathbb{R}^{N_c\times N_c}$. Notice that $\widetilde G^<_{LD}$ is a tall matrix containing only one nonzero block $G^<_{DL} \in \mathbb{C}^{N_c\times N_c}$ at the top, i.e., $\widetilde G^<_{DL}(\mathcal{I}_L,:) = G^<_{DL}$. Likewise, $\widetilde G^<_{DR}(\mathcal{I}_R,:) = G^<_{DR}$ has only one nonzero block at the bottom. The outer blocks in (\ref{eq:q}) are defined as
\begin{align}\label{eq:Glessblk}
    \widetilde G^<_{DL}(E) &= G(E)\Gamma_R(E)G(E)^{\dag}{\widetilde \tau}_{DL}(E)g_{LL}(E)^\dag \\ \nonumber
    \widetilde G^<_{DR}(E) &= G(E)\Gamma_R(E)G(E)^{\dag}{\widetilde \tau}_{DR}(E)g_{RR}(E)^\dag + G(E){\widetilde \tau}_{DR}(E)\operatorname{Im}(g_{RR}(E))
\end{align}
Again $\widetilde \tau_{DL}$ and $\widetilde \tau_{DR}$ are tall matrices with ${\widetilde \tau}_{DL}(\mathcal{I}_L,:) = \tau_{DL}$ and ${\widetilde \tau}_{DR}(\mathcal{I}_R,:) = \tau_{DR}$.

The integral (\ref{eq:Q}) is commonly evaluated by numerical integration
\begin{equation}\label{eq:quadrature}
 {Q_{neq}} =  - \frac{2}{\pi }\sum\limits_i^{{N_E}} {q\left( {{E_i}} \right){w_i}}
\end{equation}
where $N_E$ evaluation points are selected within $\left[E_{b},E_{t}\right]$ according to a specific quadrature rule, such as the Gauss Legendre quadrature, and $w_i$ are the corresponding weights (the scalar Fermi function is also incorporated into $w_i$ and omitted throughout the paper).

Computing the diagonals of the matrix products in (\ref{eq:q}) is the most time-consuming step for large problems, and several approaches have been proposed to accelerate this step. RGF is the most prevailing one wherein the computational grid is partitioned into layers along one direction (usually the transport direction), and $M$ becomes a symmetric block tri-diagonal matrix. Then any block in $G$ can be computed by a recursive algorithm using the tri-diagonal blocks of $M$. The known problems with RGF include the quasi-1D assumption and the cubically growing computation with the block size. The FIND and SelInv methods are more advanced alternatives to RGF, by partitioning the grid points into arbitrarily shaped clusters organized in a binary tree~\cite{Hetmaniuk:13}. That way, the minimum block size is no longer limited by the layer size, and can be made much smaller to enjoy computational benefits. Nevertheless, the methods mentioned above are all exact methods that do not allow tradeoff between accuracy and performance, and they rely on sparse block LU factorization which may have scalability issue when applied to truly large problems. In the next section, we will propose a new approximate formulation for NEQ density calculation which is based on solving sparse linear systems directly.

\section{A New Formulation of NEQ Charge Density based on Solution of Sparse Linear Systems}\label{sec:X-formulation}
The new formulation begins with processing the $\Gamma_R$ matrix in (\ref{eq:Gless}), which contains only one $N_c\times N_c$ nonzero block $\Gamma^{blk}_R$ at the lower right corner. We assume real $E$ to make $\tau_{DR}$ a real matrix and plug it in (\ref{eq:Sigmablk}), then
\begin{equation}\label{eq:Gammablk}
    \Gamma^{blk}_R = \operatorname{Im}(\Sigma^{blk}_R) =  \tau_{DR}\operatorname{Im}(g_{RR})\tau_{DR}^\dag
\end{equation}
We exploit the fact that $\Gamma_R^{blk}$ is real symmetric, and thus admits a symmetric Schur decomposition
\[\Gamma_R^{blk} = \widetilde{U}\widetilde{D}\widetilde{U}^\dag\]
where $\widetilde U$ is a unitary matrix and $\widetilde D$ a diagonal matrix containing the eigenvalues\footnote{The absolute values of the eigenvalues are the singular values in this case} of $\Gamma_R^{blk}$. A truncation is then performed to drop the eigenvalues smaller than a prescribed threshold (rank tolerance $\epsilon_{rank}$) and the corresponding columns in $\widetilde U$. Suppose $p$ eigenvalues are kept, the rank-$p$ approximation of $\Gamma_R^{blk}$ is
\begin{equation}\label{eq:Gamma_blk_ls}
    \Gamma_R^{blk}(E) \approx U(E)D(E)U(E)^\dag
\end{equation}
where $U \in \mathbb{R}^{N\times p}$ has unitary columns, $U^\dag U = I$, and the rank-$p$ approximation of $\Gamma_R$ reads
\begin{equation}\label{eq:Gamma_ls}
    \Gamma_R(E) \approx Y(E)D(E)Y(E)^\dag, Y(E) = \left[ {\begin{array}{*{20}{c}}
{{{\rm O}}}  \\
   U(E)  \\
 \end{array} } \right]
\end{equation}
with ${\rm O}$ being an $(N - {N_c})\times p$ zero matrix. The assumption here is that we can drop some less important information contained in the self-energy matrix and still obtain a good approximation to the $\Gamma$ to be used in charge density calculations.

Substituting (\ref{eq:Gamma_ls}) into (\ref{eq:Gless}) and (\ref{eq:Glessblk}), we obtain
\begin{align}\label{eq:Gless_ls}
    G^<(E) &\approx \left[G(E)Y(E)\right]D(E) \left[G(E)Y(E)\right]^{\dag} \\ \nonumber
    \widetilde G^<_{LD}(E) &\approx \left[G(E)Y(E)\right]D(E) \left[G(E)Y(E)\right]^{\dag}{\widetilde \tau}_{DL}(E)g_{LL}(E)^\dag \\ \nonumber
    \widetilde G^<_{DR}(E) &\approx \left[G(E)Y(E)\right]D(E) \left[G(E)Y(E)\right]^{\dag}{\widetilde \tau}_{DR}(E)g_{RR}(E)^\dag \\ \nonumber
                           & + G(E){\widetilde \tau}_{DR}(E)\operatorname{Im}(g_{RR}(E))
\end{align}
The key quantity to compute in (\ref{eq:Gless_ls}) is the solution of the following sparse linear system
\begin{equation}\label{eq:X}
    X(E) = G(E)Y(E) = \left[ES-H-\Sigma_L(E)-\Sigma_R(E)\right]^{-1}Y(E)
\end{equation}
With $X^L = X(\mathcal{I}_L,:), X^R = X(\mathcal{I}_R,:)$ denoting the top and bottom $N_c \times p$ blocks of $X$, (\ref{eq:Gless_ls}) becomes
\begin{align}\label{eq:Gless_X}
    G^<(E) &\approx X(E)D(E)X(E)^{\dag} \\ \nonumber
    G^<_{DL}(E) &\approx X^L(E)D(E)X^L(E)^{\dag}\tau_{DL}(E)g_{LL}(E)^\dag \\ \nonumber
    G^<_{DR}(E) &\approx X^R(E)D(E)X^R(E)^{\dag}\tau_{DR}(E)g_{RR}(E)^\dag + X^R(E)\beta(E)
\end{align}
where $G^<_{DL}$ and $G^<_{DR}$ are the nonzero blocks of $\widetilde G^<_{DL}$ and $\widetilde G^<_{DR}$, respectively. The $\beta$ in the last equation is defined by
\begin{equation}\label{eq:beta}
\beta(E) = U(E)^\dag\tau_{DR}(E)\operatorname{Im}(g_{RR}(E))
\end{equation}
the derivation of which is discussed in~\ref{sec:appendix1}. Equations (\ref{eq:Gless_X}) are the main results of the new formulation, which is named the {\emph{X-formulation}} for its base on the solution $X$'s of a series of sparse linear systems with block RHS. Note that if no low-rank approximation is applied to $\Gamma_R$ and (\ref{eq:X}) is solved exactly, the X-formulation becomes an exact method.

After $X$ is obtained, the diagonals of matrix products in (\ref{eq:q}), in the generic form of $diag\{X_1X_2^\dag\}$, need to be extracted. Computing the diagonals of the outer product of two $N\times p$ matrices can be done in a lower complexity than the brute-force $O(N^2p)$. For instance, $diag\left\{G^< S\right\}$ can be computed by (under Matlab notations)
\begin{equation}\label{eq:diag}
    diag\left\{G^< S\right\} = {\text{sum}}(G^<.\ast S^T,2) = {\text{sum}}(X.\ast (DX^\dag S)^T,2)
\end{equation}
where $.\ast$ represents the element-wise multiplication and ${\text{sum}}(...,2)$ denotes the summation along rows. The complexity of (\ref{eq:diag}) is reduced to $O(Np)$. Non-diagonal entries of the lesser Green's function can also be readily obtained once $X$ is available. The flow to compute the electron density $Q$ with the X-formulation is summarized in Algorithm~\ref{alg1}.
\begin{algorithm}
\DontPrintSemicolon
\LinesNumbered
\Begin{
$Q \leftarrow 0$\;
\For{$j \leftarrow 1$ \KwTo $N_E$}{
Obtain $U(E_j),D(E_j)$ via low-rank approximation (\ref{eq:Gamma_blk_ls})\;
$\beta_j \leftarrow U(E_j)^\dag\tau_{DR}(E_j)\operatorname{Im}(g_{RR}(E_j))$\;
$D_j \leftarrow w_jD(E_j)$, $\beta_j \leftarrow w_j\beta(E_j)$\;
$M_j \leftarrow E_jS-H-\Sigma(E_j)$\;
$X_j \leftarrow M_j^{-1}U(E_j)$\;
$X_{j}^L \leftarrow X_j(\mathcal{I}_L,:)$ and $X_{j}^R \leftarrow X_j(\mathcal{I}_R,:)$\;
$Q \leftarrow Q+{\text sum}\left(X_j.*(D_jX_j^\dag S)^T,2\right)$\;
$Q(\mathcal{I}_L) \leftarrow Q(\mathcal{I}_L)- {\text sum}\left(X_{j}^L.*(D_jX_{j}^{L\dag}\tau_{DL}(E_j)g_{LL}(E_j)^\dag s_{DL}^T)^T,2\right)$\;
$Q(\mathcal{I}_R) \leftarrow Q(\mathcal{I}_R)-{{sum}}\left(X_{j}^R.*((D_jX_{j}^{R\dag}\tau_{DR}(E_j)g_{RR}(E_j)^\dag + \beta_j)s_{DR}^T)^T,2\right)$\;
}
$Q \leftarrow -\frac{2}{\pi}Q$\;
}
\caption{Charge density calculation with X-formulation\label{alg1}}
\end{algorithm}

Several notes are important at this point:
\begin{enumerate}
  \item The X-formulation uses the low-rank factors of $\Gamma$ as the RHS when solving the linear systems. Therefore, the number of RHS vectors, i.e., the rank used in the low-rank approximation, is important for the performance, and serves as an adjustable parameter to cater for different accuracy and speed requirements in different applications. In typical two-terminal devices without scattering, only one nonzero block in $\Gamma$ needs to be factorized, and its size (the size of the contact with higher chemical potential) is generally small compared to whole structure. When scattering mechanisms are included, $\Gamma$ tends to become block (or tri-block) diagonal, and the X-formulation remains applicable with a low-rank approximation applied to the whole $\Gamma$, but may need to work with an increased number of RHS vectors. In addition, the low-rank approximation is performed after the self-energy matrix is formed by an exact method in this work. It would be interesting to investigate the possibility to approximate self-energy matrices in low-rank factor form in the first place using iterative methods such as~\cite{Sorensen:08}.
  \item Both direct and iterative methods can be used to solve the sparse linear systems in (\ref{eq:X}), rendering more flexibility in the X-formulation than in RGF and its variants. When direct methods are used, the X-formulation possesses a comparable complexity with FIND and SelInv as they are all based on the LDLT factorization of complex symmetric matrices. On the other hand, state-of-the-art iterative solvers such as GMRES~\cite{GMRES} and COCG~\cite{COCG} are readily applicable, which may be the only viable option for extremely large structures in term of time and memory complexity. In addition, with a block RHS and a form closed to shifted linear system (if $\Sigma$ is linear in $E$), the solutions of the equation systems (\ref{eq:X}) are expected to have linearly dependent components among RHSs and energies, and as such the subspace needed to capture all the solutions can be significantly smaller than the lumped size of that required to solve one system with one RHS at a time. Iterative methods exploiting this property, for instance the shifted COCG method~\cite{sCOCG}, can be promising candidates to be used in conjunction with the X-formulation. Parallelization may also be more straightforward with the iterative methods.
  \item The X-formulation based on (\ref{eq:X}) is essentially in a simplified parametric state-space form with the time derivative term being zero. The low-rank factor of $\Gamma$ serves as the ''input'' of the system, which is reasonable in physics since the self-energies ''capsulate'' the influence of outer environment. $X$ can be viewed as the ''internal states'' and also the ''output'' needed to extract the quantities of interest, e.g., the electron density. In this regard, the X-formulation opens new avenues for introducing well-established MOR techniques into QM calculations.
\end{enumerate}

\section{A Nonlinear MOR Scheme}
The most expensive step of the X-formulation is finding the solution $X$ with (\ref{eq:X}), which has to be repeated for a prescribed set of energy points, i.e.,
\begin{equation}\label{eq:Xj}
    X_j = \left[E_j S-H-\Sigma_L(E_j)-\Sigma_R(E_j)\right]^{-1}Y(E_j) = M_j^{-1} Y_j,\;\;\; j=1,2,...,N_E
\end{equation}
When the applied bias is high ($\sim 1V$), $N_E \approx 500$ energies may be needed to approximate the oscillatory integration. Solving $X_j$ hundreds of times poses a substantial challenge even with the state-of-the-art computing resources for large-scale problems. In this section a nonlinear MOR scheme is developed to effectively reduce the number of energies where the full-size problems are solved.

\subsection{Basic Projection-based MOR scheme}
The NEQ integration has a feature that a dense sampling is applied to a relatively small interval (compared to the equilibrium case). As such, the solutions $X$'s of all energies may have substantial overlap in information and can be approximated to a reasonable extent by a smaller number of sampling points. To achieve this, we select $m \ll N_E$ energy points as the interpolation points, at which $X$ is solved exactly (or to a sufficiently high accuracy). We assume these solutions are linearly independent and thus span the subspace
\begin{equation}\label{eq:PDKS}
 \mathcal{K}_m = {\text{span}}\left\{ {M{{\left( {{E_1}} \right)}^{ - 1}}Y\left( {{E_1}} \right),M{{\left( {{E_2}} \right)}^{ - 1}}Y\left( {{E_2}} \right), \ldots ,M{{\left( {{E_{{m}}}} \right)}^{ - 1}}Y\left( {{E_{{m}}}} \right)} \right\}
\end{equation}
The subspace above is called the parameter-dependent Krylov subspace (PDKS) \cite{Zaslavsky:10}, which has nonlinear dependency on $E$ both in $M$ and $Y$ due to the nonlinear self-energies.

Let $V_m = \left[v_1,v_2,\ldots,v_m\right]\in \mathbb{R}^{N\times {\widetilde n}}$, $\widetilde n = \sum_j^m{p_j}$, be an orthonormal basis constructed from the $m$ accurate solutions above, the solution of (\ref{eq:X}) is approximated by solving a reduced-order system obtained from a projection with $V_m$
\begin{equation}\label{eq:Xm}
    X_m(E) = V_m M_m(E)^{-1} Y_m(E)
\end{equation}
in which
\begin{align}\label{eq:MmYm}
    M_m(E) &= \left(M(E)V_m\right)^\dag M(E) V_m \\ \nonumber
    Y_m(E) &= \left(M(E)V_m\right)^\dag Y(E)
\end{align}
Note that the Petrov-Galerkin condition is enforced in (\ref{eq:MmYm}) to minimize the residual over $V_m$
\begin{equation}\label{eq:Rm}
    \|R_m(E)\|_F = \|M(E)X_m(E)-Y(E) = M(E)V_m M_m(E)^{-1} Y_m(E)-Y(E)\|_F
\end{equation}
where $\|\cdot\|_F$ denotes the Frobenius norm.
\begin{algorithm}
\DontPrintSemicolon
\LinesNumbered
\KwData{$H,S,\Sigma,\mathcal{I}_{intp}$}
\KwResult{Non-equilibrium charge density $Q$}
\Begin{
\For{$m \leftarrow 1$ \KwTo $m_{intp}$}{
$j \leftarrow \mathcal{I}_{intp}(m)$\;
$M(E_j) \leftarrow E_jS-H-\Sigma(E_j)$\;
$X_m \leftarrow M(E_j)^{-1}Y(E_j)$\;
\tcp{Modified Gram-Schmidt orthogonalization}
\eIf{$m == 1$}{
$V_1 \leftarrow qr(X_1)$\;}{$V_m \leftarrow orth([V_{m-1},X_m])$\;}
}
$W \leftarrow 0$, $W_{DL} \leftarrow 0$ and $W_{DR} \leftarrow 0$\;
\For{$j \leftarrow 1$ \KwTo $N_E$}{
$\beta_j \leftarrow U(E_j)^\dag\tau_{DR}(E_j)\operatorname{Im}(g_{RR}(E_j))$\;
$D_j \leftarrow w_jD(E_j)$, $\beta_j \leftarrow w_j\beta(E_j)$\;
Construct $M_{m}(E_j)$ and $Y_m(E_j)$ via (\ref{eq:MmYm})\;
Solve the reduced system $Z_m(E_j) \leftarrow M_m(E_j)^{-1}Y_m(E_j)$\;
$X_{j}^L \leftarrow V_m(\mathcal{I}_L,:)Z_m(E_j)$ and $X_{j}^R \leftarrow V_m(\mathcal{I}_R,:)Z_m(E_j)$\;
$W \leftarrow W+Z_m(E_j)D_jZ_m(E_j)^\dag$\;
$W_{DL} \leftarrow W_{DL}+X_{j}^L\left(D_jX_{j}^{L\dag} \tau_{DL}(E_j)g_{LL}(E_j)^\dag \right)$\;
$W_{DR} \leftarrow W_{DR}+X_{j}^R\left(D_jX_{j}^{R\dag} \tau_{DR}(E_j)g_{RR}(E_j)^\dag + \beta_j\right)$\;
}
$Q \leftarrow -\frac{2}{\pi}{\text sum}\left(V_mW.*(V_m^\dag S)^T,2\right)$\;
$Q(\mathcal{I}_L) \leftarrow \rho(\mathcal{I}_L)-(-\frac{2}{\pi}){\text sum}\left(W_{DL}.*s_{DL},2\right)$\;
$Q(\mathcal{I}_R) \leftarrow \rho(\mathcal{I}_R)-(-\frac{2}{\pi}){\text sum}\left(W_{DR}.*s_{DR},2\right)$\;
}
\caption{MOR with predetermined interpolation points\label{alg2}}
\end{algorithm}

The basic MOR method with a predetermined set of interpolation points is given in Algorithm~\ref{alg2}. The vector $\mathcal{I}_{intp}$ stores the $m_{intp}$ selected interpolation points. The modified Gram-Schmidt scheme is employed to orthogonalize $V_m$. Note that in Line $20-22$, the multiplications with the $S$ matrices are done only {\emph{one time}} after the second {\emph{for}} loop, which is much more efficient than having $S$ multiplied with $X$ every step within the {\emph{for}} loop as in the non-MOR Algorithm~\ref{alg1}. This is because $V_m$ is energy independent, and the E-dependent matrices $Z_mDZ_m^\dag \in \mathbb{C}^{\widetilde n \times \widetilde n}$ are generally of small sizes and can be formed and added to $W$ efficiently within the loop. In contrast, it is prohibitive to evaluate and store $XDX^\dag$ of the original size during the {\emph{for}} loop in Algorithm~\ref{alg1} \footnote{Although it is possible to compute and store only the entries in $XDX^\dag$ that will be needed in subsequent multiplication with $S$, the computational cost remains much higher than the evaluation of $Z_mDZ_m^\dag$}, and thus the multiplications with the $S$ matrices must be performed immediately before relevant information is lost.

The PDKS in (\ref{eq:PDKS} is generated by adding $p_j$ basis vectors at a time, where $p_j$ is the number of columns of $U(E_j)$. Therefore a linear system with $p_j$ RHSs needs to be solved each time, which may still be expensive for large $p_j$. On the other hand, not all RHS vectors have equal contribution to the error reduction, and it is natural to ask whether we can select the most important RHS vectors based on their effectiveness in reducing error. In addition, although choosing interpolation points {\emph{a priori}} is convenient, in practice determining them with high quality is difficult, if not impossible, for general nonlinear problems. In the following section, adaptive approaches are developed to optimize the selection of the RHS vectors and the interpolation points in the MOR algorithm.

\subsection{Adaptive Selection of RHS Vectors amd Interpolation Points}\label{sec:adaptive}
Firstly, to generate the RHS matrix that is cost-effective in error reduction, we extend the tangential interpolation idea applied in the rational Krylov subspace~\cite{Druskin:14} to the parametric case. The tangential interpolation refers to that, instead of solving $M_j^{-1}Y_j$ with $p_j$ RHS columns, we solve $M_j^{-1}Y_jd_j$, with $d_j \in \mathbb{C}^{p_j\times l_j},\;l_j \leq p_j$ being a ``selector matrix'' to cut the number of RHSs from $p_j$ to $l_j$. In other words, the interpolation is not exact, but along some tangential direction $d_j$, at the point $j$. The tangential PDKS thus becomes
\begin{equation}\label{eq:TPDKS}
 \mathcal{T}_m = {\text{span}}\left\{ {M{{\left( {{E_1}} \right)}^{ - 1}}Y\left( {{E_1}} \right)d_1,M{{\left( {{E_2}} \right)}^{ - 1}}Y\left( {{E_2}} \right)d_2, \ldots ,M{{\left( {{E_{{m}}}} \right)}^{ - 1}}Y\left( {{E_{{m}}}} \right)d_m} \right\}
\end{equation}

Following~\cite{Druskin:14}, we obtain $d_j$ from the singular value decomposition (SVD) of the residual $R_{j-1}(E_j)$. Specifically, $d_j$ is chosen as the right singular vectors corresponding to the $l_j$ largest singular values to maximize $\|R_{j-1}(E_j)d_j\|$. Our implementation determines $l_j$ according to a user-defined SVD tolerance ($\epsilon_{svd}$), such that
\begin{equation}\label{eq:svd}
 \sigma_{l_j} < \epsilon_{svd}\sigma_{1}
\end{equation}
where $\sigma_{1},\sigma_{2},...$ are the singular values of $R_{j-1}(E_j)$ in the descending order. Note that this tangential reduction is performed on top of the low-rank approximation of $\Gamma$ (\ref{eq:Gamma_ls}) used by the X-formulation, and thus the full-size solution in the interpolation stage of MOR involves generally fewer RHSs than in the X-formulation at the same energy.

The second issue concerns the selection of the interpolation points. More interpolation points help produce high-quality ROMs to benefit the subsequent integration, but at the price of solving more systems of the original size. For more regular problems wherein $M(E)$ are exactly shifted linear systems, e.g., $M(E) = A-EB$, priori schemes exist for choosing the interpolation points (or shifts) of optimal asymptotic convergence by solving the third Zolotaryov problem~\cite{Druskin:09}. However, such schemes are not applicable in this case due to the nonlinear, non-analytic, self-energy terms. An alternate scheme was proposed in~\cite{Druskin:10}, which chooses the interpolation points, within a given spectral interval, in a greedy fashion to minimize the residuals of the approximated systems. In this paper, we extend the greedy search approach to select the next interpolation point based on the residual estimate of the current approximation.

Given $V_m$ the projection basis generated at the $m$th step, an optimal implementation is to compute the residual estimates via (\ref{eq:Rm}) for all the unselected points, among which the one with the largest residual is chosen as the interpolation point at the $m+1$th step. However, each residual evaluation with (\ref{eq:Rm}) involves forming the dense tall matrix $MV_m$, computing the inner product to generate $M_m$ (\ref{eq:MmYm}) and solving a dense matrix equation $M_m^{-1}Y_m$. When the dimension of $M_m$ is large, checking residuals for hundreds of points may become very expensive. To speed up the point selection procedure, following strategies are adopted in our implementation:
\begin{enumerate}
  \item The interpolation process starts with a set of $m_{init}>1$ initial interpolation points $\mathcal{I}_{init}$, chosen equidistantly in the indices. After each residual sweeping, $m_{new} > 1$ points $\mathcal{I}_{new}$ with the largest residuals are selected and appended to the interpolation point vector $\mathcal{I}_{intp}$. A new check of residuals is employed only after these $m_{new}$ points are solved, and thus the {\emph{number of times to perform residual checking}} is largely reduced. However, this is a suboptimal scheme. To see this, suppose $\mathcal{I}_{new} = [j_1,j_2,...,j_{m_{new}}]$ with descending residuals are picked, then adding new basis vectors from the solution at $E_{j_1}$ will affect the residuals at all the remaining $E$'s, in particular it tends to suppress most the residuals in the vicinity of $E_{j_1}$. Thus there is no guarantee that $E_{j_2},...,E_{j_{m_{new}}}$ remain meaningful choices for the subsequent steps, if they are closed to $E_{j_1}$ and already have their residuals reduced by the solution at $E_{j_1}$. In other words, we hope that $E_{j_2},...,E_{j_{m_{new}}}$ would still be, or at least close to, the ones that would be selected in the one-point-at-a-time strategy in later stage. To this end, we force the candidate points in $\mathcal{I}_{new}$ to be separated by a minimal distance $d_{min}$. This way, the interpolation points are more likely to be allocated to the needed regions.
  \item The second strategy is to reduce the {\emph{number of points participating in a residual checking}}. Remind that in (\ref{eq:MmYm} the Petrov-Galerkin projection is applied, which is chosen deliberately over the standard Galkerkin projection to ensure the residuals drop monotonically with the expansion of projection basis. Consequently, one can safely exclude the points that have residuals already smaller than a given residual tolerance ($\epsilon_{res}$) in the residual sweep as the residuals can only become smaller later on, and fewer points need to be checked at later stage of the algorithm when the residual evaluation becomes more expensive.
  \item Finally, the {\emph{individual residual evaluation}} in (\ref{eq:Rm}) can be sped up by an incremental update of $R_m$ reusing the results from previous steps. In~\ref{sec:appendix2}, we present an efficient update scheme for $R_m$ based on incremental LDLT factorization, which dramatically reduces the computation by only working on the newly generated data in each step.
\end{enumerate}
The adaptive algorithm for choosing RHS and interpolation points is given in~\ref{alg3}.
\begin{algorithm}
\DontPrintSemicolon
\LinesNumbered
\KwData{$H,S,\Sigma,m_{max},\mathcal{I}_{init}$}
\KwResult{Projection basis $V_m$}
\Begin{
$\mathcal{I}_{intp} \leftarrow \mathcal{I}_{init}$, $m_{intp} \leftarrow m_{init}$\;
$\mathcal{I}_{remain} \leftarrow \mathcal{I}_{all}\backslash \mathcal{I}_{intp}$, $m_{remain} \leftarrow N_E-m_{intp}$\;
\For{$m \leftarrow 1$ \KwTo $m_{max}$}{
\eIf{$m \leq m_{intp}$}{
$j \leftarrow \mathcal{I}_{intp}(m)$\;
}{
\For(\tcp*[h]{Estimate residuals}){$k \in \mathcal{I}_{remain}$}{
Compute $RES(k) = \|R_m(E_k)\|_F$ via (\ref{eq:Rm}) using $V_{m-1}$\;
}
\eIf(){$\max(RES) \leq \epsilon_{res}$}{Exit \;}{
Select $m_{new}$ interpolation points $\mathcal{I}_{new}$ based on certain strategy\;
$\mathcal{I}_{intp} \leftarrow \left[\mathcal{I}_{intp},\mathcal{I}_{new}\right]$ \tcp*[h]{update $\mathcal{I}_{intp}$}\;
$m_{intp} \leftarrow m_{intp}+m_{new}$\;
$j \leftarrow \mathcal{I}_{intp}(m)$\;
Find $\mathcal{I}_{small}$ where $RES(\mathcal{I}_{small}) < \epsilon_{res}$\;
$RES(\mathcal{I}_{small}) \leftarrow 0$\;
$\mathcal{I}_{remain} \leftarrow \mathcal{I}_{remain}\backslash \left[\mathcal{I}_{new},\mathcal{I}_{small}\right]$\tcp*[f]{\small{No need to check residuals at these points}}\;
}
}
\tcp{Point $j$ will be the next interpolation point}
$M(E_j) \leftarrow E_jS-H-\Sigma(E_j)$\;
Compute the leading right singular vectors of $R_m(E_j)$ to define $d_{j}$\;
$Y_m = \left[0,U(E_j)d_j\right]^T$\;
$X_m \leftarrow M(E_j)^{-1}Y_m$\;
\eIf{$m == 1$}{
$V_1 \leftarrow qr(X_1)$\;}{$V_m \leftarrow orth([V_{m-1},X_m])$\;}
}
}
\caption{Adaptive selection of RHS vectors and interpolation points\label{alg3}}
\end{algorithm}

There are several tolerances serving for different purposes in the X-formulation $+$ MOR framework. The rank tolerance $\epsilon_{rank}$ controls the number of RHSs in solving linear systems in the X-formulation: higher $\epsilon_{rank}$ means fewer RHSs but lower accuracy in approximating the impact of self-energies. The number of RHSs also affects the time needed to perform matrix-matrix multiplications in the second {\emph{for}} loop in Algorithm~\ref{alg2}. The SVD tolerance $\epsilon_{svd}$ determines the number of RHSs solved in the interpolation phase of MOR, generally smaller than the one selected by $\epsilon_{rank}$ in the X-formulation. Smaller $\epsilon_{svd}$ tends to add more vectors to the projection basis $V$ at a time and is likely to reduce the number of interpolation points needed to achieve the same accuracy, yet requires more computation time and may result in larger ROMs. The residual tolerance $\epsilon_{res}$ directly controls the balance between accuracy and efficiency of the MOR method.

Finally, The MOR method presented above can be viewed as one particular way to reuse the information shared among the solutions at different energies, as discussed in Section~\ref{sec:X-formulation}, but by no means is the only way or the most efficient one. Investigations for more effective strategies, e.g., smarter choices of interpolation points, are highly desired to design faster algorithms for QM calculations. In addition, the current MOR method works with a given set of quadrature points determined {\emph{a priori}} by a quadrature rule. It would be interesting to incorporate the MOR scheme into the adaptive integration technique such as~\cite{Baumgartner:07} to gain further saving.

\section{Numerical Experiments}
\begin{figure}[ht]
  \centering
  \includegraphics[width=50mm]{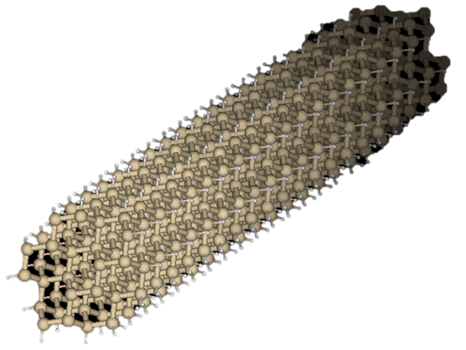}\\
  \caption{3D picture of SiNW2.}\label{fig:SNW3D}
\end{figure}

In this section we test the proposed computation framework using three 3D silicon nanowires (SiNW) of different sizes. Table~\ref{tab:size} details the specifications of the three SiNWs and Fig.~\ref{fig:SNW3D} plots the 3D structure of SiNw2. The coding is done in Matlab and the tests were performed on a machine with $2.7$GHz CPU and $32$Gb memory. To simplify the discussion, in this paper we use the direct solver PARDISO~\cite{PARDISO} to solve all the linear systems, yet iterative solvers are readily applicable if better performance can be achieved. A small imaginary part of $10^{-5}$ is added to all the energies to avoid singularity.
\begin{table}[ht]
  \centering
  \caption{Sizes of silicon nanowires}
    \begin{tabular}{cccccc}
    \toprule
    case  & cross sec. (nm$^2$) & length (nm) & \# of atoms & matrix size & contact size $N_c$ \\ \midrule
    SiNW1     & 1.5$\times$1.5  & 50    & 3,264 & 15,252 & 166 \\
    SiNW2     & 5$\times$2.5  & 25    & 12,540 & 53,268 & 1,244\\
    SiNW3     & 8$\times$8  & 20    & 54,800 & 203,000 & 4,060\\ \bottomrule
    \end{tabular}%
  \label{tab:size}%
\end{table}%

\subsection{Performance of X-Formulation}
\begin{table}[ht]
\small
  \centering
  \caption{Performance of X-formulation ($\epsilon_{rank} = 10^{-6}$) compared with RGF}
    \begin{tabular}{c|cccccc}
    \toprule
    Case  & Method & $N_x$ in RGF & \# of RHS & runtime (s) & memory (Mb) & error \\ \midrule
    \multirow{2}[0]{*}{SiNW1} & RGF   & 166   & $-$   & 2.6   & 325   & $-$ \\
          & X-formula & $-$   & 23    & 1.1   & 200   & 2.0E-07 \\ \midrule
    \multirow{2}[0]{*}{SiNW2} & RGF   & 1,244 & $-$   & 294   & 4,900 & $-$ \\
          & X-formula & $-$   & 201   & 32    & 1,350 & 4.8E-07 \\ \midrule
    \multirow{2}[0]{*}{SiNW3} & RGF   & 4,060 & $-$   & 17,334 & 51,000 & $-$ \\
          & X-formula & $-$   & 1,038 & 1,071 & 7,200 & 4.2E-07 \\ \bottomrule
    \end{tabular}%
    \normalsize
  \label{tab:X_perf}%
\end{table}
We start with comparing the performance of the X-formulation against RGF for single-energy calculation with the three examples in Table~\ref{tab:X_perf}. The absolute rank tolerance $\epsilon_{rank}$ for approximating $\Gamma_R$ is set to be $10^{-6}$ and the error is measured by $\|q-q_{RGF}\|_F/\|q_{RGF}\|_F$. All the solutions are obtained at $E=-2.1529$eV. As expected, RGF exhibits a low efficiency when the cross-section of a structure is large, with approximately a cubic increase in computation and a quadratic increase in memory with respect to $N_x$. The X-formulation, on the other hand, shows better time and memory scalability, with an over $16X$ speedup for the largest case. The accuracy is reasonable with the errors all smaller than $\epsilon_{rank}$.
\begin{figure}[ht]
  \centering
  \includegraphics[width=90mm]{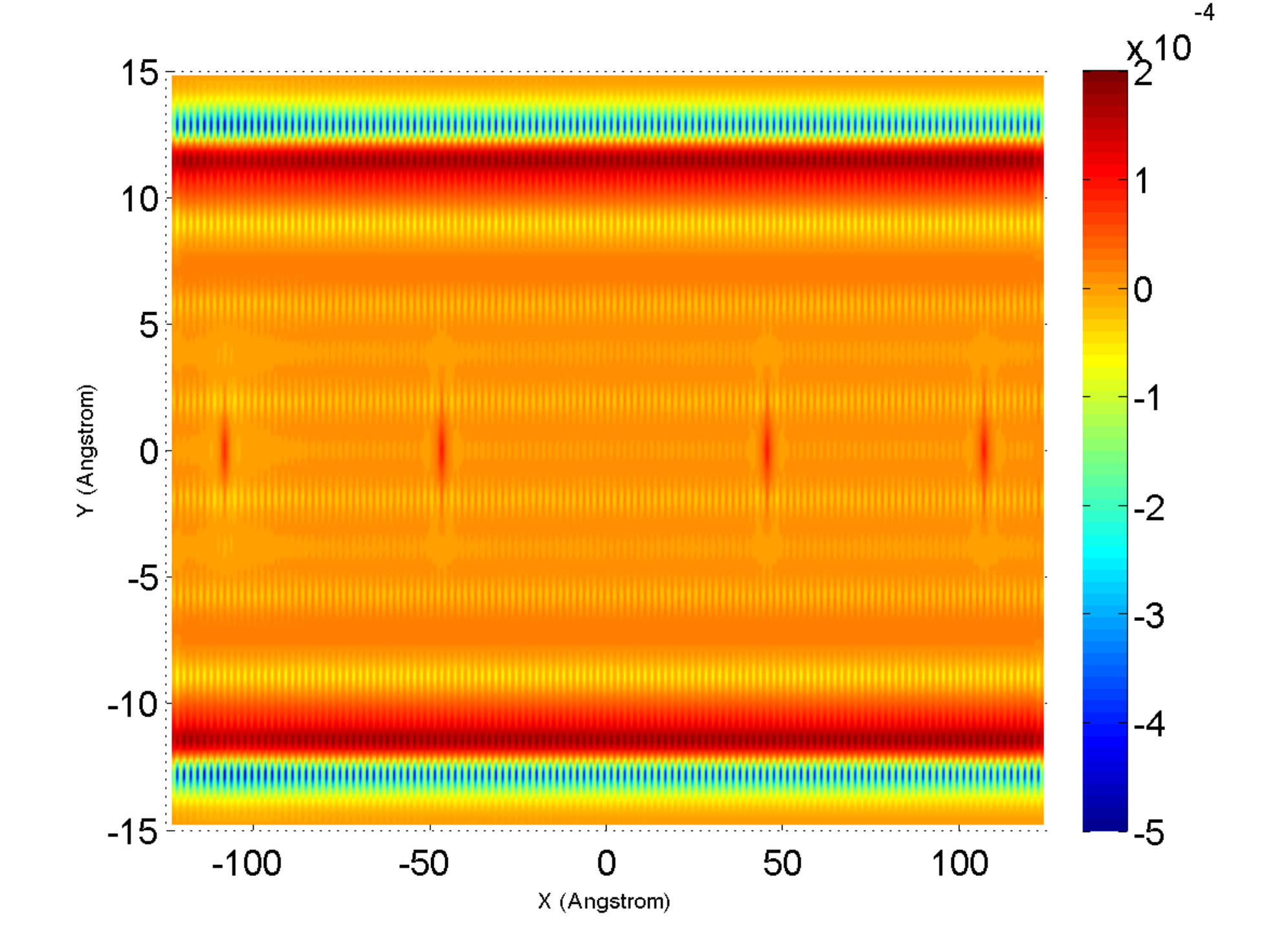}\\
  \caption{Electron density of the SiNW2 example.}\label{fig:density}
\end{figure}

\begin{figure}[ht]
  \centering
  \includegraphics[width=85mm]{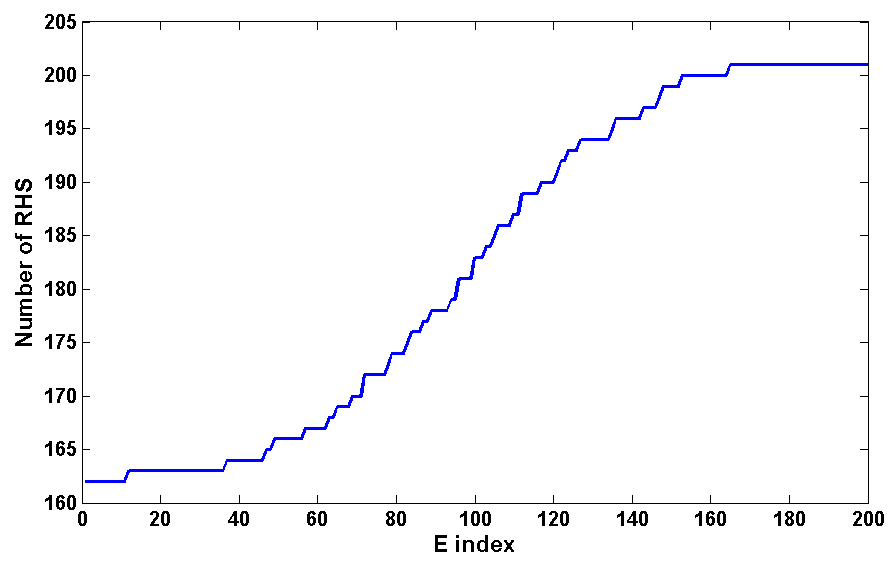}\\
  \caption{Number of RHSs at different energy points (SiNW2, $\epsilon_{rank}=10^{-6}$). }\label{fig:Nrhs}
\end{figure}

Next we compute the electron density of SiNW2 by an energy integration with $0.5eV$ bias in $[-2.652904,-2.152904]eV$, and $200$ energy points chosen by the Gauss-Legendre quadrature. The calculated electron density on the middle $x-y$ plane is plotted in Fig.~\ref{fig:density}. In Fig.~\ref{fig:Nrhs}, we show the numbers of RHSs at the $200$ points for the same $\epsilon_{rank}=10^{-6}$. As can be seen, more RHSs are kept for the energies near the right end of the spectrum, where the right lead has higher contribution to the charge density distribution in the device. Because of the low-rank compression, the RHS number is much smaller than the size of the nonzero block ($4060$) in the self-energy matrix, which is one of the main contributors for the numerical advantages of the X-formulation.

Table~\ref{tab:X_ranktol} shows the performance of X-formulation for different rank tolerances with the same set of $200$ points. The number of RHSs is averaged over the $200$ points, and the total time includes the time of the PARDISO solver and the time of the subsequent matrix-matrix multiplications. Overall, the X-formulation is found to be nearly one-order faster than RGF, and allows tradeoff between accuracy and efficiency in different applications by adjusting $\epsilon_{rank}$. Larger $\epsilon_{rank}$ results in fewer RHS vectors but less accurate solution, with an error generally below the chosen $\epsilon_{rank}$. It is also observed that the performance of direct solver is not very sensitive to the number of RHSs, i.e., the runtime of PARDISO is increased by only $1.3X$ times when the amount of RHSs increases from $4$ to $102$, which can be explained by the fact that the main cost of direct solvers is on the matrix factorization and the number of RHSs only affects the runtime in the back substitution step of a lower complexity.
\begin{table}[ht]
  \centering
  \small
  \caption{X-formulation with different rank tolerances for SiNW2 with $200$ points}
    \begin{tabular}{cccccc}
    \toprule
    method & $\epsilon_{rank}$ & \# of RHS (avg.) & PARDISO time (s) & total time (s) & error \\ \midrule
    RGF   & $-$   & $-$   & $-$   & 58,845 & $-$ \\
    X-Form & $10^{-4}$ & 4     & 3,667 & 3,793 & 1.00E-04 \\
    X-Form & $10^{-5}$ & 102   & 4,856 & 5,110 & 3.60E-06 \\
    X-Form & $10^{-6}$ & 182   & 5,817 & 6,180 & 9.70E-07 \\ \bottomrule
    \end{tabular}%
    \normalsize
  \label{tab:X_ranktol}%
\end{table}%

\subsection{Performance of MOR}
\begin{figure}[ht]
  \centering
  \includegraphics[width=80mm,height=60mm]{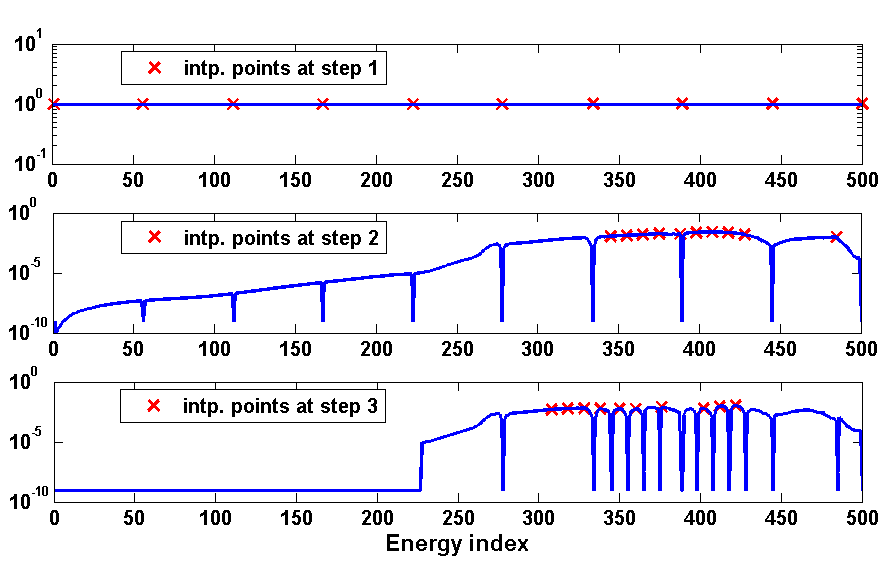}\\
  \caption{Interpolation point selection in the first three steps of MOR (SiNW1). For better visualization, the residuals at the points selected in the previous steps, together with those already smaller than $\epsilon_{res}$ (all included in $\mathcal{I}_{small}$ in Algorithm~\ref{alg3}), are all set to be $10^{-9}$.}\label{fig:point_select}
\end{figure}

We first visualize in Fig.~\ref{fig:point_select} the point selection scheme presented in Section~\ref{sec:adaptive} using SiNW1. The integration is performed in $[-2.652904,-1.652904]eV$ with $1eV$ bias and $500$ quadrature points. After each residual sweeping $m_{new} = 10$ points are selected with the minimal separation $d_{min} = 10$. Fig.~\ref{fig:point_select} shows the residuals of all energy points at the first three steps, where the interpolation points selected are marked by red crosses. In the 1st step, the points are distributed equidistantly over the whole range. In the 2nd step, the residuals at the points selected in the 1st round have been reduced small, and $10$ new points are put to the places where the residuals are the largest, subject to the constraint that each point must be separated from others by at least $d_{min}$ points. Similar selection is done in the 3rd step. Note that, after step 2, some points at the left end already have residuals smaller than $\epsilon_{rank}$, therefore these points are in the ``safe zone'' and will not participate in the residual search of step 3.

\begin{figure}[ht]
  \centering
  \includegraphics[width=85mm]{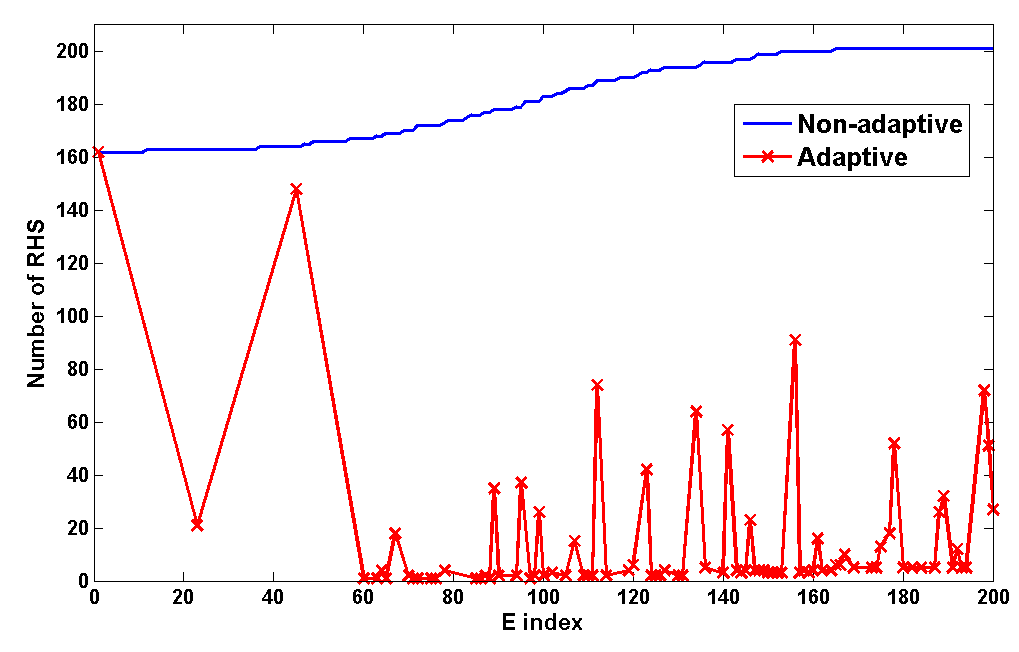}\\
  \caption{Number of RHSs without and with adaptive selection (SiNW2, $\epsilon_{rank}=10^{-6}$).}\label{fig:Nrhs_adaptive}
\end{figure}

The performance of MOR is recorded in Table~\ref{tab:perf_MOR} with SiNW1 and SiNW2 using the integration setting as mentioned above. $T_{slc}$, $T_{PAR}$ and $T_{int}$ refer to the time spent on the point selection, the PARDISO solutions and the numerical integration using the ROM, respectively. The error is measured against the reference solution obtained by solving the X-formulation directly at all the energy points. A $5X$ saving is achieved for SiNW1, wherein only $77$ full-size systems are solved instead of $500$, and the integration is performed with a ROM of size $408$ in contrast to the original size of $15,252$. The reduction in the number of full-size solutions is less significant for SiNW2, where $88$ out of $200$ energy points are required. However the runtime saving in solving linear systems, $1,669s$ vs. $6,180s$, is higher than the $2.5X$ saving in the number of energies, which can be explained by that fewer RHSs are involved in each full-size solution in the interpolation phase of MOR thanks to the adaptive selection of RHSs described in Section~\ref{sec:adaptive}. Fig.~\ref{fig:Nrhs_adaptive} compares the RHS numbers needed in the non-adaptive and the adaptive schemes. Except at the first energy the two schemes solve the same amount of RHSs, at the remaining points selected for interpolation, adaptive scheme uses a much smaller number of RHSs because the singular values of residuals drop rapidly and only a few of them deserve to be kept. This results in faster interpolation and smaller ROMs.
\begin{table}[htbp]
  \centering
  \small
  \caption{Performance of MOR. The same set of parameters are used: $\epsilon_{rank}=10^{-6}$, $\epsilon_{res}=10^{-5}$, $\epsilon_{svd}=10^{-2}$, $m_{init} = m_{new} = 10$ and $d_{min} = 10$}
    \begin{tabular}{rrccccc}
    \toprule
    case  & \# of E & \shortstack{non-MOR \\ time (s)} & \shortstack{\# of intp. \\ points} & \shortstack{ROM \\size} & \shortstack{MOR time (s) \\ ($T_{slc}+T_{PAR}+T_{int}$)} & error \\ \midrule
    SiNW1 & 500 & 660   & 77    & 408   & \shortstack{123 \\ (27+80+16)} & 2.7e-6 \\ \midrule
    SiNW2 & 200 & 6,180 & 88    & 1,337 & \shortstack{2,375 \\ (374+1,669+332)} & 1.8e-5 \\ \bottomrule
    \end{tabular}%
    \normalsize
  \label{tab:perf_MOR}%
\end{table}%

\section{Conclusion}
In this paper we present a new approximate framework for NEGF transport calculation. The two key ingredients are the X-formulation for single-energy charge density calculation based on solving sparse linear systems with block RHSs, and the MOR method to address the difficulty arising from many energy points required for large biases. The former exploits matrix sparsity directly, enabling tradeoff between accuracy and performance, and transforms the problem into a state-space form that allows MOR to be applied. The MOR method then aims to reuse the information shared among individual solutions by building a small ROM from a small fraction of energy points. A series of acceleration techniques, including the adaptive selection of RHS vectors and interpolation points and the incremental update of residual estimates, are developed to further enhance the performance of MOR. The numerical experiments confirm the advantages of the proposed framework.

\section{Acknowledgments}
The authors would like to acknowledge the support from Hong Kong University Grant Council (AoE/P-04/08) and Hong Kong General Research Fund (GRF) project HKU710913E.

\bibliography{MOR}

\appendix
\section{Formation of $\beta$}\label{sec:appendix1}
Hereafter the dependency on $E$ is dropped when it is clear from the context. The motivation to introduce $\beta$ in (\ref{eq:Gless_X}) is to reuse the same solution $X^R$, generated using $U$ as RHS, to compute the second term for $\widetilde G^<_{DR}$ in (\ref{eq:Gless_ls}). To this end, we approximate ${\tau _{DR}}\operatorname{{Im}} \left( {{g_{RR}}} \right)$ by its orthogonal projection onto $U$
\begin{equation}\label{eq:beta_proj}
{\tau _{DR}}\operatorname{{Im}} \left( {{g_{RR}}} \right) = UU^\dag {\tau _{DR}}\operatorname{{Im}} \left( {{g_{RR}}} \right)
\end{equation}
whereby $\beta$ is derived as the ''coordinate'' matrix under the basis $U$
\[\beta  = {U^\dag }{\tau _{DR}}\operatorname{{Im}} \left( {{g_{RR}}} \right)\]

It is observed in our experiments that (\ref{eq:beta_proj}) is generally a good approximation provided that $UDU^\dag$ is a good approximation of $\Gamma_R$. However, unlike the well-known error bound for SVD-based rank-$k$ approximation
\begin{equation}\label{eq:err_Gamma}
    {\left\| {{\Gamma _R} - UD{U^\dag }} \right\|_F} = \sum\limits_{i = k + 1}^{{N_c}} {{\sigma _i}},
\end{equation}
where $\sigma_i$ are the singular values sorted in the descending order, the error bound for (\ref{eq:beta_proj}) is less obvious. Hence we ensure the accuracy of (\ref{eq:beta_proj}) by a numerical means, wherein we measure the projection error by
\begin{equation}\label{eq:err_beta}
    {\left\| {{U_N}U_N^\dag {\tau _{DR}}\operatorname{Im} \left( {{g_{RR}}} \right)} \right\|_F}
\end{equation}
where $U_N$ denotes the orthogonal complement of $U$ and is available in the Schur decomposition (\ref{eq:Gammablk}). If the error is large, we move some vectors from $U_N$ to $U$ to enlarge the projection subspace to suppress the error. In our experiments fewer than $50$ vectors need to be added to $U$ in the worst case, and only the result of $Q(\mathcal{I}_R)$ will be slightly affected. One may also avoid this complication by applying another low-rank approximation to ${\tau _{DR}}\operatorname{{Im}} \left( {{g_{RR}}} \right)$ directly, and solving the additional RHS vectors thereby generated. As demonstrated in Table~\ref{tab:X_ranktol}, the increase in cost is expected to be mild since matrix factors in direct methods or preconditioners in iterative methods can be reused.

\section{Incremental Calculation of Residuals}\label{sec:appendix2}
Since only the bottom block of the RHS matrix $Y$ is nonzero, we also evaluate the residual at the same block for faster computation. This is justified in that the residual estimates need not to be exact, and the saving from a faster residual evaluation can easily outweigh the cost of solving slightly more points due to a less optimal selection scheme. Therefore the residual used in the residual sweeping reads
\begin{equation}\label{eq:Rmblk}
    R_m^{blk} = M(\mathcal{I}_R,:)V_m M_m^{-1} Y_m-U
\end{equation}
We focus on accelerating the two most expensive operations, namely, forming $M_m$ and solving $M_m^{-1}Y_m$. For the first part, we aim to build $M_m$ in an incremental manner by reusing the result from previous steps. Recall that
\begin{equation}\label{eq:Mm_update}
     M_m= {\left( {M{V_m}} \right)^\dag }M{V_m} = {\left( {M\left[ {{V_{m - 1}},{v_m}} \right]} \right)^\dag }M\left[ {{V_{m - 1}},{v_m}} \right] = \left[ {\begin{array}{*{20}{c}}
   {{M_{11}}} & {{M_{12}}}  \\
   {M_{12}^\dag } & {{M_{22}}}  \\
 \end{array} } \right]
\end{equation}
where $V_{m-1}$ is the aggregated basis vectors from the first $m-1$ steps, and $v_m$ contains the new vectors generated at the $m$th step. The blocks in $M_m$ are given by
\begin{equation}\label{Mm_blk}
\begin{gathered}
  {M_{11}} = {M_{m - 1}} = {\left( {M{V_{m - 1}}} \right)^\dag }M{V_{m - 1}} \hfill \\
  {M_{12}} = {\left( {M{V_{m - 1}}} \right)^\dag }M{v_m} \hfill \\
  {M_{22}} = {\left( {M{v_m}} \right)^\dag }M{v_m} \hfill \\
\end{gathered}
\end{equation}
in which $M_{m-1}$ is the old matrix from the last step, and $M_{12}$ and $M_{22}$ are the new blocks we need to compute. In addition, $M_{12}$ and $M_{22}$ can be obtained efficiently by minimizing the amount of computation involving $E$-dependent self-energy matrices. Take $M_{12}$ for instance
\begin{equation}\label{eq:M12}
\small
    \begin{gathered}
  {\left( {M{V_{m - 1}}} \right)^\dag }M{v_m} = {\left( {ES{V_{m - 1}} - H{V_{m - 1}} - \Sigma {V_{m - 1}}} \right)^\dag }\left( {ES{v_m} - H{v_m} - \Sigma {v_m}} \right) = {A_1} - {A_2} \hfill \\
  {A_1} = {\left| E \right|^2}{\left( {S{V_{m - 1}}} \right)^\dag }S{v_m} - \bar E{\left( {S{V_{m - 1}}} \right)^\dag }H{v_m} - E{\left( {S{v_m}} \right)^\dag }H{V_{m - 1}} + {\left( {H{V_{m - 1}}} \right)^\dag }H{v_m} \hfill \\
  {A_2} = {\left( {ES{V_{m - 1}} - H{V_{m - 1}} - \frac{1}
{2}\Sigma {V_{m - 1}}} \right)^\dag }\left( {\Sigma {v_m}} \right) + {\left( {\Sigma {V_{m - 1}}} \right)^\dag }\left( {ES{v_m} - H{v_m} - \frac{1}
{2}\Sigma {v_m}} \right) \hfill \\
\end{gathered}
\normalsize
\end{equation}
The $A_1$ term does not involve $\Sigma$ and is linear in $E$, thus one can pre-compute the $E$-independent matrices priori to the residual checking and perform only scalar-matrix multiplications within the loop. Meanwhile $SV_{m},HV_{m}$ can be expanded incrementally in the same fashion as in (\ref{eq:Mm_update}). The nonlinear $A_2$ term has to be evaluated for each $E$. However, both $\Sigma {V_{m - 1}}$ and $\Sigma {v_m}$ have only two nonzero blocks (the top and bottom blocks), so it suffices to compute only the corresponding blocks in the other two terms, which reduces the whole computation to four block matrix-matrix multiplications of relatively small sizes. Again, one can grow $\Sigma {V_{m}}$ incrementally by storing the two nonzero blocks. The efficient computation of $M_{22}$ follows analogously.

For $M_m^{-1}Y_m$, we apply the block LU-update~\cite{Huynh:08} to generate the block factorization of $M_m$, which reduces to block LDLT-update in this case since $M_m$ is Hermitian. Let $M_{11}$ in (\ref{eq:Mm_update}) have the LDLT factorization $ldl\left(M_{11}\right)=L_{11}D_{11}L_{11}^\dag$\footnote{In practice a permutation matrix $P$ is usually needed to make $L$ truly low-triangular, i.e., $M_{11}=P_{11}L_{11}D_{11}(P_{11}L_{11})^\dag$, but its incorporation is straightforward and can be updated in the same way as $D$}, then the block factorization of the augmented matrix can be updated as
\begin{equation}\label{eq:LDLT_update}
    \left[ {\begin{array}{*{20}{c}}
   {{M_{11}}} & {{M_{12}}}  \\
   {M_{12}^\dag } & {{M_{22}}}  \\
 \end{array} } \right] = \left[ {\begin{array}{*{20}{c}}
   {{L_{11}}} & {}  \\
   {{L_{21}}} & {{L_{22}}}  \\
 \end{array} } \right]\left[ {\begin{array}{*{20}{c}}
   {{D_{11}}} & {}  \\
   {} & {{D_{22}}}  \\
 \end{array} } \right]{\left[ {\begin{array}{*{20}{c}}
   {{L_{11}}} & {}  \\
   {{L_{21}}} & {{L_{22}}}  \\
 \end{array} } \right]^\dag }
\end{equation}
where
\begin{equation}\label{eq:L21L22}
    L_{21}^\dag  = D_{11}^{ - 1}L_{11}^{ - 1}{M_{12}},\;\;\;{L_{22}} = ldl\left( {{M_{22}} - {L_{21}}{D_{11}}L_{21}^\dag } \right)
\end{equation}
In other words, if $L_{11},D_{11}$ are stored from the previous step, the new $L,D$ factors can be obtained easily by one back substitution and one LDLT of a small matrix with the newly generated $M_{12}$ and $M_{22}$.

Despite being fast, the incremental residual update has a drawback that it needs to store the intermediate matrices, such as $\Sigma {V_{m}}$, for all the energies remaining in the residual checklist. Strategies to lower memory usage, e.g., the low-rank tensor approximation~\cite{Kolda:09} given the 3-dimensional data structure, will be investigated in the future.

\end{document}